\documentstyle[12pt,aps]{revtex}

\widetext
\draft
\tighten
\begin{document}

\preprint{EFUAZ FT-96-30}

\title{Different Quantum Field Constructs in the
$(1/2,0)\oplus (0,1/2)$ Representation\thanks{Submitted to ``Phys. Lett. B".}}

\author{{\bf Valeri V. Dvoeglazov}}

\address {
Escuela de F\'{\i}sica, Universidad Aut\'onoma de Zacatecas \\
Antonio Doval\'{\i} Jaime\, s/n, Zacatecas 98068, ZAC., M\'exico\\
Internet address:  VALERI@CANTERA.REDUAZ.MX
}

\date{July 25, 1996}

\maketitle

\bigskip

\begin{abstract}
We present another concrete realization of a quantum field theory,
envisaged many years ago by Bargmann, Wightman and Wigner.
Considering the special case of the $(1/2,0)\oplus (0,1/2)$
field and developing the Majorana-McLennan-Case-Ahluwalia construct
for neutrino we show that fermion and its antifermion can have same
intrinsic parities.  The construct can be applied to explanation of the
present situation in neutrino physics.
\end{abstract}


\newpage

While thirty years passed since the proposal of the Glashow-Weinberg-Salam
model, we are still far from understanding  many its essential theoretical
ingredients; first of all, fundamental origins of ``parity violation"
effect, the Kobayshi-Maskawa mixing and Higgs phenomenon. Experimental
neutrino physics and astrophysics provided us by new puzzles, that until
now did not find adequate explanation. For instance, recently Prof.
Bilenky~\cite{Bil} pointed out that it follows from the analysis of the LSND
neutrino oscillation signal that ``there is no natural hierarchy
of coupling among generations  in the lepton sector". Moreover,
at the same time the atmospheric neutrino anomaly indicates at ``the
existence of an additional sterile neutrino state besides the three active
flavor neutrino states".

The Majorana idea~\cite{Majorana}, recently analyzed in detail by
Ahluwalia~\cite{DVA94}, gives alternative way of describing
neutral particles, which is based on the treatment of self/anti-self
charge conjugate states. This formalism is believed at the moment to
be able to provide a natural mechanism of neutrino oscillations through
the Majorana mass term in the Lagrangian.

In ref.~\cite{DVA94} in the framework of the Majorana-McLennan-Case
kinematical scheme the following type-II bispinors of the $(j,0)\oplus
(0,j)$ representation space have been defined in the momentum
representation:
\begin{eqnarray} \lambda(p^\mu)\,\equiv \,\pmatrix{ \left
( \zeta_\lambda\,\Theta_{[j]}\right )\,\phi^\ast_{_L}(p^\mu)\cr
\phi_{_L}(p^\mu)} \,\,,\quad \rho(p^\mu)\,\equiv \, \pmatrix{
\phi_{_R}(p^\mu)\cr
\left ( \zeta_\rho\,\Theta_{[j]}\right )^\ast
\phi^\ast_{_R}(p^\mu)} \,\,\quad .\label{sp-dva}
\end{eqnarray}
$\zeta_\lambda$ and $\zeta_\rho$ are the phase factors that
are  fixed by the conditions of  self/anti-self charge conjugacy,
$\Theta_{[j]}$ is the Wigner time-reversal operator for spin $j$.
In the present essay we show that the construct based on the type-II
spinors  leads to another example of the
Nigam-Foldy-Bargmann-Wightman-Wigner (FNBWW) type quantum field
theory.

The irreducible projective  representations of the
quantum-mechanical Poincar\`e group have been enumerated by
Wigner~\cite{W1,W2}. He showed that one has to distinguish four
cases. The Dirac field, that describes the eigenstates
of the charge operator, belongs to
the simplest one.\footnote{Nevertheless, let us still not forget
that the Dirac construct allows one to describe both
particle and its antiparticle which have opposite eigenvalues
of the charge operator.} In three other ones there is a phenomenon which
could be called as doubling of an ordinary Fock space (or, in the
Schr\"odinger language, doubling the number of components of the wave
function). An explicit example of the FNBWW-type quantum field theory
has recently been presented~\cite{DVA00} in the $(1,0)\oplus (0,1)$
representation of the extended Lorentz group (see also earlier
papers~\cite{Nigam,Gelfand,Sila}). The remarkable feature of the construct
presented in~\cite{DVA00} is the fact that in such a framework a boson and
its antiboson have opposite intrinsic parities. In this letter
we present a construct in which fermion and antifermion have same
intrinsic parities. We prove this by working out explicitly their
properties under operators of discrete symmetries $C$, $P$ and $T$.

Let us begin with the transformation properties of the left $\phi_{_L}$
(and $\chi_{_L} = (\zeta_\rho^\ast \Theta_{[j]}) \phi_{_R}^\ast$),  and
the right $\phi_{_R}$ (and $\chi_{_R} = (\zeta_\lambda \Theta_{[j]})
\phi_{_L}^\ast$) 2-spinors.  In particular, the $(1/2,0)$ spinors
transform with respect to restricted Lorentz transformations according to
the Wigner's rules:
\begin{mathletters} \begin{eqnarray}
\phi_{_R} (p^\mu) &=& \Lambda_{_R} (p^\mu \leftarrow
\overcirc{p}^\mu)\phi_{_R} (\overcirc{p}^\mu)\, = \,\exp (+
\,\frac{\bbox{\sigma}\cdot \bbox{\varphi}}{2}) \phi_{_R}
(\overcirc{p}^\mu)\quad,\label{b01}\\
\chi_{_R} (p^\mu) &=& \Lambda_{_R} (p^\mu
 \leftarrow \overcirc{p}^\mu) \chi_{_R} (\overcirc{p}^\mu) \,= \, \exp (+
\,\frac{\bbox{\sigma}\cdot \bbox{\varphi}}{2}) \chi_{_R}
(\overcirc{p}^\mu)\quad,\label{b02}
\end{eqnarray}
\end{mathletters}
and the $(0,1/2)$ spinors,
\begin{mathletters} \begin{eqnarray}
\phi_{_L} (p^\mu) &=& \Lambda_{_L} (p^\mu \leftarrow \overcirc{p}^\mu)
\phi_{_L} (\overcirc{p}^\mu) \,=\, \exp (- \,\frac{\bbox{\sigma}\cdot
\bbox{\varphi}}{2}) \phi_{_L} (\overcirc{p}^\mu) \quad,\label{b1}\\
\chi_{_L} (p^\mu) &=& \Lambda_{_L} (p^\mu \leftarrow
\overcirc{p}^\mu)\chi_{_L} (\overcirc{p}^\mu) \,=\, \exp (-
\,\frac{\bbox{\sigma}\cdot \bbox{\varphi}}{2}) \chi_{_L}
(\overcirc{p}^\mu)\quad,\label{b2} \end{eqnarray} \end{mathletters} where
$\bbox{\varphi}$ are the Lorentz boost parameters, {\it
e.g.}~\cite{Ryder}, $\bbox{\sigma}$ are the Pauli matrices.  In the chiral
representation one can choose the spinorial basis (zero-momentum spinors)
in the following way:\footnote{Overall phase factors of left- and right-
spinors are assumed to be the same, see formulas (22a,b) in ref.~[3c]. In
this paper we try to keep the notation of the cited reference.}
\begin{mathletters}
\begin{eqnarray}
\lambda^S_\uparrow (\overcirc{p}^\mu)
&=& \sqrt{{m\over 2}}\pmatrix{0\cr i\cr 1\cr 0}\, ,\,
\lambda^S_\downarrow (\overcirc{p}^\mu) =
\sqrt{{m\over 2}}\pmatrix{-i\cr 0\cr 0\cr 1}\, ,\,
\lambda^A_\uparrow (\overcirc{p}^\mu)
= \sqrt{{m\over 2}}\pmatrix{0\cr -i\cr 1\cr 0}\, ,\,
\lambda^A_\downarrow (\overcirc{p}^\mu) = \sqrt{{m\over 2}}
\pmatrix{i\cr 0\cr 0\cr 1}\, ,\,\\
\rho^S_\uparrow (\overcirc{p}^\mu)
&=& \sqrt{{m\over 2}}\pmatrix{1\cr 0\cr 0\cr -i}\, ,\,
\rho^S_\downarrow (\overcirc{p}^\mu)
= \sqrt{{m\over 2}}\pmatrix{0\cr 1\cr i\cr 0}\, ,\,
\rho^A_\uparrow (\overcirc{p}^\mu)
= \sqrt{{m\over 2}}\pmatrix{1\cr 0\cr 0\cr i}\, ,\,
\rho^A_\downarrow (\overcirc{p}^\mu)
= \sqrt{{m\over 2}}\pmatrix{0\cr 1\cr -i\cr 0}\, .
\end{eqnarray}
\end{mathletters}
The indices $\uparrow\downarrow$ should be referred to the chiral helicity
quantum number introduced in ref.~\cite{DVA94}.
Using the boost (\ref{b01}-\ref{b2}) the reader would immediately
find the 4-spinors of the second kind  $\lambda^{S,A}_{\uparrow\downarrow}
(p^\mu)$ and $\rho^{S,A}_{\uparrow\downarrow} (p^\mu)$
in an arbitrary frame:
\begin{mathletters}
\begin{eqnarray}
\lambda^S_\uparrow (p^\mu) &=& \frac{1}{2\sqrt{E+m}}
\pmatrix{ip_l\cr i (p^- +m)\cr p^- +m\cr -p_r}\quad,\quad
\lambda^S_\downarrow (p^\mu)= \frac{1}{2\sqrt{E+m}}
\pmatrix{-i (p^+ +m)\cr -ip_r\cr -p_l\cr (p^+ +m)}\quad,\quad\\
\lambda^A_\uparrow (p^\mu) &=& \frac{1}{2\sqrt{E+m}}
\pmatrix{-ip_l\cr -i(p^- +m)\cr (p^- +m)\cr -p_r}\quad,\quad
\lambda^A_\downarrow (p^\mu) = \frac{1}{2\sqrt{E+m}}
\pmatrix{i(p^+ +m)\cr ip_r\cr -p_l\cr (p^+ +m)}\quad,\quad\\
\rho^S_\uparrow (p^\mu) &=& \frac{1}{2\sqrt{E+m}}
\pmatrix{p^+ +m\cr p_r\cr ip_l\cr -i(p^+ +m)}\quad,\quad
\rho^S_\downarrow (p^\mu) = \frac{1}{2\sqrt{E+m}}
\pmatrix{p_l\cr (p^- +m)\cr i(p^- +m)\cr -ip_r}\quad,\quad\\
\rho^A_\uparrow (p^\mu) &=& \frac{1}{2\sqrt{E+m}}
\pmatrix{p^+ +m\cr p_r\cr -ip_l\cr i (p^+ +m)}\quad,\quad
\rho^A_\downarrow (p^\mu) = \frac{1}{2\sqrt{E+m}}
\pmatrix{p_l\cr (p^- +m)\cr -i(p^- +m)\cr ip_r}\quad.
\end{eqnarray}
\end{mathletters}
with $p_r = p_x + i p_y$, $p_l = p_x -ip_y$, $p^{\pm} = p_0 \pm p_z$.
Therefore, one has~[Eqs.(48a,48b),3c]
\begin{mathletters}
\begin{eqnarray}
\rho^S_\uparrow (p^\mu) \,&=&\, - i \lambda^A_\downarrow (p^\mu)\quad,\quad
\rho^S_\downarrow (p^\mu) \,=\, + i \lambda^A_\uparrow (p^\mu)\quad,\quad\\
\rho^A_\uparrow (p^\mu) \,&=&\, + i \lambda^S_\downarrow (p^\mu)\quad,\quad
\rho^A_\downarrow (p^\mu) \,=\, - i \lambda^S_\uparrow (p^\mu)\quad.
\end{eqnarray}
\end{mathletters}
The normalization of the spinors $\lambda^{S,A}_{\uparrow\downarrow}
(p^\mu)$ and $\rho^{S,A}_{\uparrow\downarrow} (p^\mu)$ are as follows:
\begin{mathletters}
\begin{eqnarray}
\overline \lambda^S_\uparrow (p^\mu) \lambda^S_\downarrow (p^\mu) \,&=&\,
- i m \quad,\quad
\overline \lambda^S_\downarrow (p^\mu) \lambda^S_\uparrow (p^\mu) \,= \,
+ i m \quad,\quad\\
\overline \lambda^A_\uparrow (p^\mu) \lambda^A_\downarrow (p^\mu) \,&=&\,
+ i m \quad,\quad
\overline \lambda^A_\downarrow (p^\mu) \lambda^A_\uparrow (p^\mu) \,=\,
- i m \quad,\quad\\
\overline \rho^S_\uparrow (p^\mu) \rho^S_\downarrow (p^\mu) \, &=&  \,
+ i m\quad,\quad
\overline \rho^S_\downarrow (p^\mu) \rho^S_\uparrow (p^\mu)  \, =  \,
- i m\quad,\quad\\
\overline \rho^A_\uparrow (p^\mu) \rho^A_\downarrow (p^\mu)  \,&=&\,
- i m\quad,\quad
\overline \rho^A_\downarrow (p^\mu) \rho^A_\uparrow (p^\mu) \,=\,
+ i m\quad.
\end{eqnarray}
\end{mathletters}
All other conditions are equal to zero (provided that
$\vartheta^{_{L,R}}_1 + \vartheta^{_{L,R}}_2 = \pi$).

First of all,  one must deduce equations for the
Majorana-like spinors in order to see what dynamics do the neutral
particles have. It is obvious that the equations (30,31) of the cited
reference~[3c] are  hard to be suitable for building the Lagrangian
dynamics (they are very unwieldy).
Nevertheless, one can use another generalized form of the Ryder-Burgard
relation (cf. Eq. (26) of~[3c] and ref.~\cite{DVO95}) for zero-momentum
spinors:
\begin{equation}\label{rbug12} \left [\phi_{_L}^h
(\overcirc{p}^\mu)\right ]^* = (-1)^{1/2-h}\, e^{-i(\vartheta_1^L
+\vartheta_2^L)} \,\Theta_{[1/2]} \,\phi_{_L}^{-h} (\overcirc{p}^\mu)\quad,
\end{equation}
Relations for zero-momentum right spinors are obtained with the
substitution $L \leftrightarrow R$. $h$ is the helicity quantum number for
the left- and right 2-spinors. Hence, implying that $\lambda^S (p^\mu)$
(and $\rho^A (p^\mu)$) answer for positive-frequency solutions; $\lambda^A
(p^\mu)$ (and $\rho^S (p^\mu)$), for negative-frequency solutions, one can
deduce the dynamical coordinate-space equations~[11c]
\begin{mathletters}
\begin{eqnarray}
i \gamma^\mu \partial_\mu \lambda^S (x) - m \rho^A (x) &=& 0 \quad,
\label{11}\\
i \gamma^\mu \partial_\mu \rho^A (x) - m \lambda^S (x) &=& 0 \quad,
\label{12}\\
i \gamma^\mu \partial_\mu \lambda^A (x) + m \rho^S (x) &=& 0\quad,
\label{13}\\
i \gamma^\mu \partial_\mu \rho^S (x) + m \lambda^A (x) &=& 0\quad.
\label{14}
\end{eqnarray}
\end{mathletters}
They can be written in the 8-component form as follows:
\begin{mathletters}
\begin{eqnarray}
\left [i \Gamma^\mu \partial_\mu - m\right ] \Psi_{_{(+)}} (x) &=& 0\quad,
\label{psi1}\\
\left [i \Gamma^\mu \partial_\mu + m\right ] \Psi_{_{(-)}} (x) &=& 0\quad,
\label{psi2}
\end{eqnarray}
\end{mathletters}
with
\begin{eqnarray}
\Psi_{(+)} (x) = \pmatrix{\rho^A (x)\cr
\lambda^S (x)\cr}\, ,\,
\Psi_{(-)} (x) = \pmatrix{\rho^S (x)\cr
\lambda^A (x)\cr}\, ,\quad \mbox{and}\quad
\Gamma^\mu =\pmatrix{0 & \gamma^\mu\cr
\gamma^\mu & 0\cr}\quad.
\end{eqnarray}
One can also re-write the equations into the two-component form.
Similar formulations have been presented by M. Markov~\cite{Markov}, and
A. Barut and G. Ziino~\cite{Ziino}.

The Dirac-like and Majorana-like field operators can
be built from both $\lambda^{S,A} (p^\mu)$ and $\rho^{S,A} (p^\mu)$,
or their combinations (see formulas Eqs. (46,47,49) in ref.~[3c]). For
instance,
\begin{equation}
\Psi (x^\mu) \equiv \int {d^3 {\bf p}\over (2\pi)^3} {1\over 2E_p}
\sum_\eta \left [ \lambda^S_\eta (p^\mu) \, a_\eta ({\bf p}) \,\exp
(-ip\cdot x) +\lambda^A_\eta (p^\mu)\, b^\dagger_\eta ({\bf p}) \,\exp
(+ip\cdot x)\right ]\quad.\label{oper}
\end{equation}
Operators of discrete
symmetries (charge conjugation and space inversion) are given by
\begin{eqnarray}
S^c_{[1/2]} = e^{i\vartheta^c_{[1/2]}} \pmatrix{0 &
i\Theta_{[1/2]}\cr -i\Theta_{[1/2]} & 0\cr}{\cal K}={\cal C}_{[1/2]} {\cal
K}\quad,\quad S^s_{[1/2]} = e^{i\vartheta^s_{[1/2]}} \pmatrix{0 &
\openone_2\cr \openone_2 & 0\cr}= e^{i\vartheta^s_{[1/2]}} \gamma^0 \quad.
\end{eqnarray}
In the Fock space operations of the charge conjugation and space
inversions can be defined through unitary operators such that:
\begin{equation}
U^c_{[1/2]} \Psi (x^\mu) (U^c_{[1/2]})^{-1} = {\cal C}_{[1/2]}
\Psi^\dagger_{[1/2]} (x^\mu)\quad,\quad
U^s_{[1/2]} \Psi (x^\mu) (U^s_{[1/2]})^{-1} = \gamma^0
\Psi (x^{\prime^{\,\mu}})\quad,
\end{equation}
the time reversal operation, through {\it an antiunitary}
operator\footnote{Let us remind that the operator of hermitian conjugation
does not act on $c$-numbers on the left side of the equation (\ref{tr}).
This fact is conected with the properties of an antiunitary operator:
$\left [ V^{^T} \lambda A (V^{^T})^{-1}\right ]^\dagger =
\left [\lambda^\ast V^{^T} A (V^{^T})^{-1}\right ]^\dagger =
\lambda \left [ V^{^T} A^\dagger (V^{^T})^{-1} \right ]$.}
\begin{equation}
\left [V^{^T}_{[1/2]}  \Psi (x^\mu)
(V^{^T}_{[1/2]})^{-1} \right ]^\dagger = S(T) \Psi^\dagger
(x^{{\prime\prime}^\mu}) \quad,\label{tr}
\end{equation}
with
$x^{\prime^{\,\mu}} \equiv (x^0, -{\bf x})$ and $x^{{\prime\prime}^{\,\mu}}
=(-x^0,{\bf x})$.  We  further assume the vacuum state to be assigned an
even $P$- and $C$-eigenvalue and, then, proceed as in ref.~\cite{DVA00}.

As a result we have the following properties of creation (annihilation)
operators in the Fock space:
\begin{mathletters}
\begin{eqnarray}
U^s_{[1/2]} a_\uparrow ({\bf p}) (U^s_{[1/2]})^{-1} &=& - ia_\downarrow
(-  {\bf p})\quad,\quad
U^s_{[1/2]} a_\downarrow ({\bf p}) (U^s_{[1/2]})^{-1} = + ia_\uparrow
(- {\bf p})\quad,\quad\\
U^s_{[1/2]} b_\uparrow^\dagger ({\bf p}) (U^s_{[1/2]})^{-1} &=&
+ i b_\downarrow^\dagger (- {\bf p})\quad,\quad
U^s_{[1/2]} b_\downarrow^\dagger ({\bf p}) (U^s_{[1/2]})^{-1} =
- i b_\uparrow (- {\bf p})\quad,
\end{eqnarray} \end{mathletters}
what signifies that the states created by the operators $a^\dagger
({\bf p})$ and $b^\dagger ({\bf p})$ have very different properties
with respect to the space inversion operation, comparing with
Dirac states (the case also regarded in~\cite{Ziino}):
\begin{mathletters}
\begin{eqnarray}
U^s_{[1/2]} \vert {\bf p},\,\uparrow >^+ &=& + i \vert -{\bf p},\,
\downarrow >^+\quad,\quad
U^s_{[1/2]} \vert {\bf p},\,\uparrow >^- = + i
\vert -{\bf p},\, \downarrow >^-\quad,\quad\\
U^s_{[1/2]} \vert {\bf p},\,\downarrow >^+ &=& - i \vert -{\bf p},\,
\uparrow >^+\quad,\quad
U^s_{[1/2]} \vert {\bf p},\,\downarrow >^- =  - i
\vert -{\bf p},\, \uparrow >^-\quad.
\end{eqnarray}
\end{mathletters}
For the charge conjugation operation in the Fock space we have
two physically different possibilities. The first one, {\it e.g.},
\begin{mathletters}
\begin{eqnarray}
U^c_{[1/2]} a_\uparrow ({\bf p}) (U^c_{[1/2]})^{-1} &=& + b_\uparrow
({\bf p})\quad,\quad
U^c_{[1/2]} a_\downarrow ({\bf p}) (U^c_{[1/2]})^{-1} = + b_\downarrow
({\bf p})\quad,\quad\\
U^c_{[1/2]} b_\uparrow^\dagger ({\bf p}) (U^c_{[1/2]})^{-1} &=&
-a_\uparrow^\dagger ({\bf p})\quad,\quad
U^c_{[1/2]} b_\downarrow^\dagger ({\bf p})
(U^c_{[1/2]})^{-1} = -a_\downarrow^\dagger ({\bf p})\quad,
\end{eqnarray}
\end{mathletters}
in fact, has some similarities with the Dirac construct.
The action of this operator on the physical states are
\begin{mathletters}
\begin{eqnarray}
U^c_{[1/2]} \vert {\bf p}, \, \uparrow >^+ &=& + \,\vert {\bf p},\,
\uparrow >^- \quad,\quad
U^c_{[1/2]} \vert {\bf p}, \, \downarrow >^+ = + \, \vert {\bf p},\,
\downarrow >^- \quad,\quad\\
U^c_{[1/2]} \vert {\bf p}, \, \uparrow >^-
&=&  - \, \vert {\bf p},\, \uparrow >^+ \quad,\quad
U^c_{[1/2]} \vert
{\bf p}, \, \downarrow >^- = - \, \vert {\bf p},\, \downarrow >^+ \quad.
\end{eqnarray} \end{mathletters}
But, one can also construct the charge conjugation operator in the
Fock space which acts, {\it e.g.}, in the following manner:
\begin{mathletters}
\begin{eqnarray}
\widetilde U^c_{[1/2]} a_\uparrow ({\bf p}) (\widetilde U^c_{[1/2]})^{-1}
&=& - b_\downarrow ({\bf p})\quad,\quad \widetilde U^c_{[1/2]}
a_\downarrow ({\bf p}) (\widetilde U^c_{[1/2]})^{-1} = - b_\uparrow
({\bf p})\quad,\quad\\
\widetilde U^c_{[1/2]} b_\uparrow^\dagger ({\bf p})
(\widetilde U^c_{[1/2]})^{-1} &=& + a_\downarrow^\dagger ({\bf
p})\quad,\quad
\widetilde U^c_{[1/2]} b_\downarrow^\dagger ({\bf p})
(\widetilde U^c_{[1/2]})^{-1} = + a_\uparrow^\dagger ({\bf p})\quad,
\end{eqnarray}
\end{mathletters}
and, therefore,
\begin{mathletters}
\begin{eqnarray}
\widetilde U^c_{[1/2]} \vert {\bf p}, \, \uparrow >^+ &=& - \,\vert {\bf
p},\, \downarrow >^- \quad,\quad
\widetilde U^c_{[1/2]} \vert {\bf p}, \, \downarrow
>^+ = - \, \vert {\bf p},\, \uparrow >^- \quad,\quad\\
\widetilde U^c_{[1/2]} \vert
{\bf p}, \, \uparrow >^- &=& + \, \vert {\bf p},\, \downarrow >^+
\quad,\quad
\widetilde U^c_{[1/2]} \vert {\bf p}, \, \downarrow >^- = + \, \vert {\bf
p},\, \uparrow >^+ \quad.
\end{eqnarray}
\end{mathletters}
Investigations of several important cases, which are different from the
above ones, are required a separate paper to. Next, by straightforward
verification one can convince ourselves about correctness of the
assertions made in~\cite{DVA94,DVA96} (see also~\cite{Nigam}) that it is
possible a situation when the operators of the space inversion and
charge conjugation commute each other in the Fock space. For instance,
\begin{mathletters}
\begin{eqnarray}
U^c_{[1/2]} U^s_{[1/2]} \vert {\bf
p},\, \uparrow >^+ &=& + i U^c_{[1/2]}\vert -{\bf p},\, \downarrow >^+ =
+ i \vert -{\bf p},\, \downarrow >^- \quad,\\
U^s_{[1/2]} U^c_{[1/2]} \vert {\bf
p},\, \uparrow >^+ &=& U^s_{[1/2]}\vert {\bf p},\, \uparrow >^- = + i
\vert -{\bf p},\, \downarrow >^- \quad.
\end{eqnarray} \end{mathletters}
The second choice of the charge conjugation operator answers for the case
when the $\widetilde U^c_{[1/2]}$ and $U^s_{[1/2]}$ operations
anticommute:
\begin{mathletters} \begin{eqnarray}
\widetilde U^c_{[1/2]} U^s_{[1/2]} \vert {\bf p},\, \uparrow >^+ &=&
+ i \widetilde U^c_{[1/2]}\vert -{\bf
p},\, \downarrow >^+ = -i \, \vert -{\bf p},\, \uparrow >^- \quad,\\
U^s_{[1/2]} \widetilde U^c_{[1/2]} \vert {\bf p},\, \uparrow >^+ &=& -
U^s_{[1/2]}\vert {\bf p},\, \downarrow >^- = + i \, \vert -{\bf p},\,
\uparrow >^- \quad.
\end{eqnarray} \end{mathletters}

Next, one can compose states which would have somewhat similar
propertiesto those which we have become accustomed.
The states $\vert {\bf p}, \,\uparrow >^+ \pm
i\vert {\bf p},\, \downarrow >^+$ answer for positive (negative) parity,
respectively.  But, what is important, {\it the antiparticle states}
(moving backward in time) have the same properties with respect to the
operation of space inversion as the corresponding {\it particle states}
(as opposed to $j=1/2$ Dirac particles).  This is again in accordance with
the analysis of Nigam and Foldy, and Ahluwalia.  The states which are
eigenstates of the charge conjugation operator in the Fock space are
\begin{equation}
U^c_{[1/2]} \left ( \vert {\bf p},\, \uparrow >^+ \pm i\,
\vert {\bf p},\, \uparrow >^- \right ) = \mp i\,  \left ( \vert {\bf p},\,
\uparrow >^+ \pm i\, \vert {\bf p},\, \uparrow >^- \right ) \quad.
\end{equation}
There is no a simultaneous set of states which were ``eigenstates" of the
operator of the space inversion and of the charge conjugation
$U^c_{[1/2]}$.

Finally, the time reversal {\it anti-unitary} operator in
the Fock space should be defined in such a way the formalism to be
compatible with the $CPT$ theorem. If we wish the Dirac states to transform
as $V(T) \vert {\bf p}, \pm 1/2 > = \pm \,\vert -{\bf p}, \mp 1/2 >$ we
have to choose (within a phase factor), ref.~\cite{Itzy}:
\begin{equation}
S(T) = \pmatrix{\Theta_{[1/2]} &0\cr 0 &
\Theta_{[1/2]}\cr}\quad.
\end{equation}
Thus, in the first relevant case we obtain for the $\Psi
(x^\mu)$ field, Eq.  (\ref{oper}):\footnote{In connection with the
proposal of the eight-component equation we still note that some
modifications in arguments concerning the formalism for time-reversal
operation are possible.}
\begin{mathletters} \begin{eqnarray}
V^{^T} a^\dagger_\uparrow ({\bf p}) (V^{^T})^{-1} &=& a^\dagger_\downarrow
(-{\bf p})\quad,\quad
V^{^T} a^\dagger_\downarrow ({\bf p}) (V^{^T})^{-1} = -
a^\dagger_\uparrow (-{\bf p})\quad,\quad \\
V^{^T} b_\uparrow ({\bf p}) (V^{^T})^{-1} &=& b_\downarrow
(-{\bf p})\quad,\quad
V^{^T} b_\downarrow ({\bf p}) (V^{^T})^{-1} = -
b_\uparrow (-{\bf p})
\end{eqnarray} \end{mathletters}

To summarise we note that  we have constructed another explicit example of
the Bargmann-Wightman-Wigner theory. The matters of  physical dynamics
connected with this mathematical construct should be solved in future as
depended on what gauge interactions with potential fields do we
introduce~[11c] and what experimental setup do we choose.

\acknowledgments
It is my pleasure to thank Prof. D. V. Ahluwalia and Prof. A. F. Pashkov
for helpful advice.   The papers of Profs.  M.  A.  Markov, I. M. Gelfand
and M. L. Tsetlin, G.  A. Sokolik, B.  P.  Nigam and L.  L. Foldy, A. O.
Barut and G. Ziino helped me to overcome old barriers and to see new
perspectives which I searched formerly.

I am grateful to Zacatecas University for a professorship.
This work has been partly
supported by el Mexican Sistema Nacional de Investigadores, el
Programa de Apoyo a la Carrera Docente, and  by the CONACyT under the
research project 0270P-E.

\end{document}